\documentclass[12pt,pre,aps,preprint,showpacs]{revtex4}

\usepackage{graphicx}
\usepackage{amsfonts}
\usepackage{amsmath}
\usepackage{mathtools}
\usepackage{float}
\usepackage{epsfig}

\newcommand{\e}[1]{\ensuremath{\times 10^{#1}}}
\newcommand{\EnthalpyPerParticle}{h}
\newcommand{\NbrPerCell}{{n}}

\newcommand{\VolumePerParticle}{v}
\newcommand{\AllLatticeVectors}{\mathbf A}

\begin{document}

\title{Inverse Statistical Mechanics: Probing the Limitations of Isotropic Pair Potentials to Produce Ground-State Structural Extremes}

\author{G. Zhang}

\email{gezhang@princeton.edu}

\affiliation{\emph{Department of Chemistry}, \emph{Princeton University},
Princeton NJ 08544 }

\author{F. H. Stillinger}

\email{fhs@princeton.edu}

\affiliation{\emph{Department of Chemistry}, \emph{Princeton University},
Princeton NJ 08544}

\author{S. Torquato}

\email{torquato@electron.princeton.edu}

\affiliation{\emph{Department of Chemistry, Department of Physics,
Princeton Institute for the Science and Technology of
Materials, and Program in Applied and Computational Mathematics}, \emph{Princeton University},
Princeton NJ 08544}


\begin{abstract}
Inverse statistical-mechanical methods have recently been employed to design optimized short-ranged radial (isotropic) pair potentials that robustly produce novel targeted classical ground-state many-particle configurations. 
The target structures considered in those studies were low-coordinated crystals with a high degree of symmetry. 
In this paper, we further test the fundamental limitations of radial pair potentials by targeting crystal structures with appreciably less symmetry, 
including those in which the particles have different local structural environments. 
These challenging target configurations demanded that we modify previous inverse optimization techniques. 
In particular, we first find local minima of a candidate enthalpy surface
and determine the enthalpy difference $\Delta H$ between such inherent structures and the target structure. 
Then we determine the lowest positive eigenvalue $\lambda_0$ of the Hessian matrix of the enthalpy surface at the target configuration. 
Finally, we maximize $\lambda_0 \Delta H$ so that the target structure is both locally stable and globally stable with respect to the inherent structures. 
Using this modified optimization technique, we have designed short-ranged radial pair potentials that stabilize the two-dimensional kagome crystal, the rectangular kagome crystal, and rectangular lattices, as well as the three-dimensional structure of CaF$_2$ crystal inhabited by a single particle species. 
We verify our results by cooling liquid configurations to absolute zero temperature via simulated annealing and ensuring that such states have stable phonon spectra.
Except for the rectangular kagome structure, all of the target structures can be stabilized with monotonic repulsive potentials. 
Our work demonstrates that single-component systems with short-ranged radial pair potentials can counterintuitively self-assemble into crystal ground states with low symmetry and different local structural environments.
Finally, we present general principles that offer guidance in determining whether certain target structures can be achieved as ground states by radial pair potentials.

\end{abstract}

\maketitle

\section{Introduction}
A fundamental problem of statistical mechanics is the determination of the phase diagram of interacting many-particle systems in the absence of an external field. 
For a single-component system of $N$ particles in a large region of volume $V$ in $d$-dimensional Euclidean space $\mathbb R^d$, 
the interaction is represented by the potential energy $\Phi(\mathbf r^N)$, 
where $\mathbf r^N=\mathbf r_1,\mathbf r_2,...,\mathbf r_N$ denotes the configurational coordinates.
A theoretically simple and computationally widely used form of the potential energy is the following pairwise form:
\begin{equation}
\Phi(\mathbf r^N)=\sum_{i<j}u_2(r_{ij}),
\end{equation}
where $u_2(r)$ is an isotropic pair potential and $r_{ij}$ is the distance between the $i$th and $j$th particles.

Even for this simple class of potentials, our understanding of the phase diagram, including the $T=0$ ground state, is still far from complete.
Two approaches have been used to study phase diagrams of isotropic pair potentials. 
In the {\it forward} approach, one first specifies the isotropic pair potential $u_2(r)$ and then probes the structures in its phase diagram.
This venerable approach has identified a variety of structures with varying degrees of complexity and order
\cite{weeks1971role, dijkstra1994phase, watzlawek1999phase, hemmer2001solid, rabani2003drying, hynninen2006prediction, noya2006structural, prestipino2009zero, zhang2010reentrant, reichhardt2010structural, batten2011novel, ji2011three, campos2012structural, zhao2012analysis}. 
In the {\it inverse} approach, a target many-particle configuration or physical property is first specified and then 
one attempts to determine an isotropic pair potential $u_2(r)$ under certain constraints that achieves the targeted behavior \cite{torquato2009inverse}.
The target behavior can be ground state configurations
\cite{rechtsman2005optimized, rechtsman2006designed, rechtsman2006self, rechtsman2007synthetic, marcotte2011optimized, marcotte2011unusual, marcotte2012designed, jain2013inverse} 
or excited-state properties, such as negative thermal expansion \cite{rechtsman2007negative} and negative Poisson ratio \cite{rechtsman2008negative}.

This paper focuses on the use of inverse statistical mechanics to determine isotropic pair potentials that produce unusual targeted crystalline structures as unique ground states, 
as in multiple previous works
\cite{rechtsman2005optimized, rechtsman2006designed, rechtsman2006self, rechtsman2007synthetic, marcotte2011optimized, marcotte2011unusual, marcotte2012designed, jain2013inverse}.
Contrary to the conventional view that low-coordinated crystal structures require directional bonds as in chemical covalency, 
earlier works employing the inverse approach have found optimized isotropic pair potentials (under certain constraints) stabilizing a variety of low-coordinated crystal structures as ground states.
Target structures that have successfully been stabilized include the square lattice \cite{rechtsman2006designed, marcotte2011optimized, marcotte2011unusual} and
honeycomb crystal \cite{rechtsman2005optimized, rechtsman2006designed, marcotte2011optimized, marcotte2011unusual} 
in two dimensions, and the simple cubic lattice \cite{rechtsman2006self, jain2013inverse}, diamond crystal \cite{rechtsman2007synthetic, marcotte2012designed, jain2013inverse}, 
and wurtzite crystal \cite{rechtsman2007synthetic} in three dimensions. 
These isotropic pair potentials have been designed using the following steps \cite{rechtsman2005optimized, rechtsman2006designed,
marcotte2011optimized, marcotte2011unusual,rechtsman2006self,rechtsman2007synthetic,marcotte2012designed, jain2013inverse}:
A functional form was chosen for the isotropic pair potential in terms of some parameters. 
One then optimized an objective function that is related to the stability of the target structure over competitors 
(for example, energy difference \cite{rechtsman2006designed} or the target structure's stable pressure range \cite{marcotte2012designed}).
Subsequently, the validity of the optimized potential was verified by cooling liquid configurations to absolute zero temperature via simulated annealing and by establishing that 
the target structure contains no phonon instabilities \cite{rechtsman2006designed}.
These results provide good counterexamples to the aforementioned intuition that low-coordinated structures require directional bonding. 
However, all of these target structures are globally highly symmetric, and the local environments around each of the particles
in these structures are identical up to spatial inversions or rotations.

Here, we further probe the limitations of isotropic pair potentials to produce ground-state structural extremes using inverse statistical-mechanical techniques.
Doing so has required us to improve upon previous optimization algorithms devised for inverse statistical mechanics for reasons that we will elaborate below.
Our improved optimization algorithm not only allows each competitor structure to deform to become more competitive during the optimization, 
but also incorporates the local mechanical stability of the target structure (i.e. enthalpy cost to deform the target structure) into our objective function.
We test our improved optimization algorithm by targeting the standard kagome crystal, 
rectangular lattices, the rectangular kagome crystal, and the three dimensional CaF$_2$ crystal inhabited by a single particle species.
Compared to previous target structures, these new targets have lower symmetry, and particles in some cases have different local structural environments.
We restrict ourselves to short-ranged potentials (i.e. $u_2(r) \equiv 0$ for $r>r_c$, where $r_c$ is a constant) 
because they are both computationally easier to treat and experimentally simpler to realize.
For all of our targets, except for the rectangular kagome crystal, we are able to stabilize them with smooth short-ranged monotonic repulsive potentials, 
which would be easier to produce experimentally.
For the rectangular kagome crystal, we found that a potential with a shallow well is needed for the class of functions considered.

In contrast to some previous inverse statistical mechanical approaches \cite{rechtsman2005optimized, rechtsman2006designed, rechtsman2006self, rechtsman2007synthetic, marcotte2011optimized, marcotte2011unusual}, 
in which the specific volume $v=V/N$ ($N$ is the number of particles and $V$ is the volume) is fixed 
and the classical ground state is achieved by the global minimum of the potential energy $\Phi(\mathbf r^N)$, 
we fix the pressure $p$ rather than the specific volume.
At constant pressure $p$ and number of particles $N$, the classical ground state is achieved by the global minimum of the configurational enthalpy per particle:
\begin{equation}
\label{enthalpy_def}
h(\mathbf r^N)=\Phi(\mathbf r^N)/N + pv.
\end{equation}
There are two advantages to fixing the pressure rather than the specific volume. 
First, at zero temperature, phase separation (coexistence) only occurs at a unique pressure for a given potential, while it can occur over a nontrivial range of densities.
By fixing the pressure rather than the density during simulations, we minimize our risk of encountering phase separation.
Second, allowing the volume to change will enable us to fully deform the simulation box, thus minimizing the boundary effect during simulations.

The rest of the paper is organized as follows: In Sec.~\ref{technique}, we describe the our improved algorithm. In Sec.~\ref{results}, we present our designed isotropic pair potentials
for the two-dimensional (2D) kagome crystal, rectangular lattices, and the rectangular kagome crystal, and the three-dimensional (3D) structure of the CaF$_2$ crystal inhabited by a single particle species. 
We close with conclusions and discussion in Sec.~\ref{conclusion}.

\section{Extended Optimization Technique}
\label{technique}
\subsection{Basic Definitions}
A {\it lattice} in $\mathbb R^d$ is an infinite periodic structure in which the space $\mathbb R^d$ is divided into identical regions called {\it fundamental cells}, 
each of which contains just one point specified by the {\it lattice vector}
\begin{equation}
\mathbf R=n_1 \mathbf a_1 + n_2 \mathbf a_2 +\dots + n_d \mathbf a_d,
\end{equation}
where $\mathbf a_i$ are the lattice vectors and $n_i$ spans all the integers for $i=1, 2, \dots,d$.
A {\it crystal} is a more general notion than a lattice because it is obtained by placing a fixed configuration of $\NbrPerCell$ points (where $\NbrPerCell \ge 1$), 
located at $\mathbf r_1, \mathbf r_2, ..., \mathbf r_\NbrPerCell$, within one fundamental cell.
The coordination structure of a crystal can be represented by the theta series $\theta$ \cite{Sloane2003Theta, conway1998packing}, 
which is the generating function of squared distances of the vector displacements between any two particles of the crystal structure and has the following form:
\begin{equation}
\label{ThetaSeries}
\theta(q)=1+ \sum_{j=1} ^ {\infty} Z_j q^{r_j^2},
\end{equation}
where $r_j$ is the distance from a particle at the origin (measured in units of the nearest neighbor distance)
and $Z_j$ is the associated average coordination number (average number of particles at a radial distance $r_j$).
See Appendix~\ref{targets_structure} for the vectors that specify the particle locations and lattice vectors of the crystal as well as the first few terms of the corresponding theta series of our target structures.
For the special case of periodic structures, Eq.~(\ref{enthalpy_def}) can be written more explicitly in terms of coordination structure:
\begin{equation}
\label{enthalpyperparticle}
\EnthalpyPerParticle(\mathbf r^\NbrPerCell; \AllLatticeVectors) =\frac{1}{2}\sum_{j} u_2(r_j)Z_j + p\VolumePerParticle(\AllLatticeVectors),
\end{equation}
where $\AllLatticeVectors=[\mathbf a_1, \mathbf a_2, ..., \mathbf a_d]^T$ is the generator matrix \cite{conway1998packing} (a matrix whose rows consist of the lattice vectors) and $\VolumePerParticle(\AllLatticeVectors)$ is the specific volume, which depends on $\AllLatticeVectors$.
The ground state is achieved by the global minimum of enthalpy per particle $\EnthalpyPerParticle(\mathbf r^\NbrPerCell; \AllLatticeVectors)$.

For each target crystal structure, we use the following steps to attempt to find an isotropic pair potential $u_2(r)$ and a pressure $p$ such that the target is the ground state.

\subsection{Search for Degenerate Ground States}
\label{DegeneraceCheck}
A target configuration cannot possibly be the unique ground state if a different structure has exactly the same coordination structure up to the range of the potential 
and the same specific volume $\VolumePerParticle$.
In this degeneracy searching step, we start from a random configuration and minimize the ``difference'' between the coordination structure of the configuration and that of the target structure;
see Appendix~\ref{coordination_difference} for a detailed description.
After minimizing the difference, if there is no difference between the two coordination structures and specific volumes, we check if the resulting configuration is equivalent to the target structure. 
Two structures are considered to be ``equivalent'' if they are related to each other through translations, rotations, inversions, uniform scalings, or combinations of the above transformations \cite{jiao2010geometrical, gommes2012microstructural}.
If the resulting configuration is different from the target structure, 
then we have found a degenerate structure and thus have proven that the target structure cannot be the unique ground state of any isotropic pair potential. 
If after trying to minimize the difference multiple times (often thousands of times) no degenerate structure is found, 
we tentatively assume that the target structure is unique and continue to the next step.
In this step, we visually inspect the configurations to determine whether two structures are equivalent. 
However, in the upcoming optimization and verification steps, since we have already assumed that the target structure has a unique coordination structure,
we can test whether another structure is equivalent to the target structure by comparing their coordination structures using the computer.

\subsection{Optimization}
\label{Optimization}
If the target structure has a unique coordination structure, it might be stabilized by an isotropic pair potential with finite range. 
We can specify a family of potential functions and optimize for the target structure's stability. 
Since extremely long-ranged potentials are both computationally inefficient and experimentally challenging to realize, 
we restrict ourselves to potential functions with compact support of the following form:
\begin{equation}
u_2(r)=
\begin{cases}
\displaystyle \left(\frac{b}{r^{12}}+c_0+c_1r+c_2r^2+... \right)\exp(-\alpha r^2)(r-r_{c})^2,& \text{if $r<r_{c}$,} \\
0,& \text{otherwise,}
\end{cases}
\label{function_form}
\end{equation}
where $b$, $c_n$ ($n=0,1,2,...$), $r_{c}$, and $\alpha$ are parameters. 
This form is realistic because it contains a stiff core $b/r^{12}$ and smoothly approaches zero as $r$ approaches $r_{c}$. 
If this form does not work well, we will add additional terms of different types, for example, Gaussian wells centered at some $r>0$.
Since the energy and length scale of the pair potential are arbitrary, we fix these scales so that:
\begin{enumerate}
\item The nearest neighbor distance of the target structure is 1.
\item The absolute value of the pair potential at the nearest neighbor distance of the target structure, $|u_2(1)|$, is 1.
\end{enumerate}
We also require that $\alpha \ge 0$ so that the effect of the Gaussian core is to decrease $u_2(r)$ as $r$ increases rather than to increase $u_2(r)$.
We further require that $r_{c} \le 6.4$ in order to ensure that the potential is relatively short-ranged.

\begin{figure}[h!]
\begin{center}
\includegraphics[width=100mm]{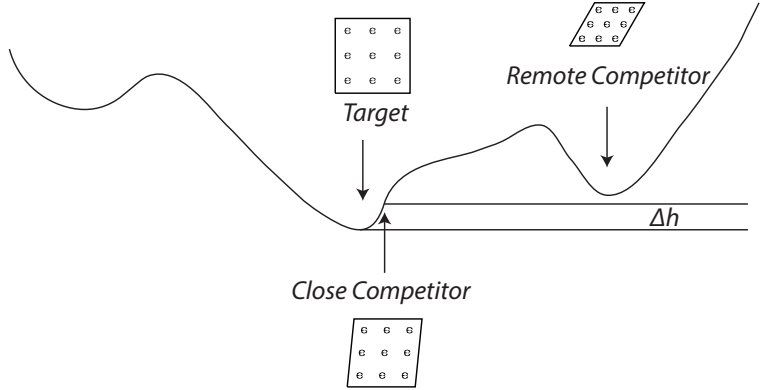}
\end{center}
\caption{A schematic plot of the enthalpy surface (equivalent of potential energy surface at constant pressure). 
If we simply define $\Delta \EnthalpyPerParticle$ as the enthalpy difference between the target and the lowest competitor and maximize it, we will encounter the ``close-competitor problem.'' 
If the competitor list contains structurally close competitors, $\Delta \EnthalpyPerParticle$ will be controlled by a structurally close competitor, causing an abnormal lifting of the enthalpy of structurally close competitors.
}
\label{CloseCompetitorProblem}
\end{figure}

After the potential form is chosen, we optimize the parameters. 
Although previous objective functions worked for previous target structures with high symmetry, they must be modified for less symmetric and more complex target structures.
The result of maximizing the energy difference or enthalpy difference is very sensitive to structurally close competitors (i.e. a slight deformation of the target structure) 
because they are not differentiated from structurally remote competitors (competitors that are not structurally close competitors).
Figure~\ref{CloseCompetitorProblem} illustrates the close-competitor problem schematically.

\begin{figure}[h!]
\begin{center}
\includegraphics[width=160mm]{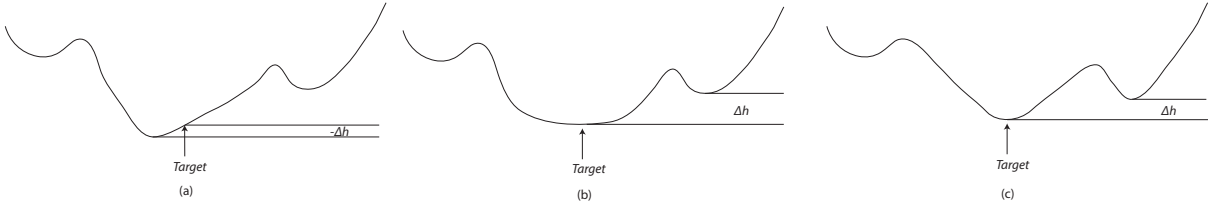}
\end{center}
\caption{A schematic plot of the enthalpy surface, illustrating our definition of $\Delta \EnthalpyPerParticle$.
(a): If the target structure is not a local minimum of the enthalpy surface, the inherent structure of the target will not be identical to the target and will have a lower enthalpy. 
$\Delta \EnthalpyPerParticle$ becomes negative.
(b) and (c): If the target structure is a local minimum of the enthalpy surface, the inherent structure of the target will be identical to the target. $\Delta \EnthalpyPerParticle$ becomes the enthalpy difference between the target and a different inherent structure. Thus $\Delta \EnthalpyPerParticle$ might be positive.
(b): However, after maximizing $\Delta \EnthalpyPerParticle$, the curvature near the target structure might be very small, leading to an undesirable phonon spectrum. 
(c): By maximizing $\lambda_0\Delta \EnthalpyPerParticle/(1+r_c^d)$, 
we sacrifice some $\Delta \EnthalpyPerParticle$ to increase the curvature near the target structure while favoring short-ranged potentials. 
Note that we usually cannot find all inherent structures in the complex, multi-dimensional enthalpy surface. 
If we miss a inherent structure that has a lower enthalpy than our target, that inherent structure will be discovered in the later verification step by simulated annealing.
}
\label{PotentialSurface_Cartoon}
\end{figure}

Optimization over a pressure range solves the close competitor problem \cite{marcotte2012designed}, but introduces its own problems. 
First, some structures with lower symmetry do not naturally have a stable pressure range. 
For example, consider the rectangular lattice with aspect ratio $b/a \ne 1$. (A precise definition of rectangular lattices and their aspect ratio is given in Appendix~\ref{targets_structure}.)
Since the structure is anisotropic, it is expected to have different elastic constants in different directions; see Appendix~\ref{ElasticConstants} for examples. 
Thus, when the pressure is changed by a small amount, the aspect ratio will also change.
Second, after the optimization, there will be many competitors that are enthalpically close to the target. 
However, these competitor structures can be very different from the target, and converting from one to another may require crossing a large enthalpy barrier. 
This makes it especially hard to find the ground state in the later simulated annealing step.

In this paper, we introduce an improved objective function that removes these shortcomings, 
enabling us to target ground-state structures with considerably greater complexity than previous targets. 
The improved objective function of the optimization is calculated by the following steps:
\begin{enumerate}
\item Given a set of potential function parameters, an isotropic pair potential $u_2(r)$ is determined. 
Using this potential function, we calculate the pressure of the target structure (knowing that the nearest neighbor distance of the target structure is 1).
For this pressure, we find the inherent structure of the target structure and each competitor structure. 
The inherent structures are obtained by minimizing the enthalpy per particle $\EnthalpyPerParticle(\mathbf r^\NbrPerCell; \AllLatticeVectors)$ in the isobaric ensemble, 
changing particle positions $\mathbf r^\NbrPerCell$ and lattice vectors $\AllLatticeVectors$.
In the current implementation, the minimizations are performed with the MINOP algorithm \cite{dennis1979two}.
\item Then, we compare each of the inherent structures with the target structure to test if they are structurally equivalent.
\item For each inherent structure that is not equivalent to the target, we calculate its enthalpy per particle $\EnthalpyPerParticle_c$. After calculating all the $\EnthalpyPerParticle_c$'s, we find their minimum value, $\EnthalpyPerParticle_{c0}$.
The difference between $\EnthalpyPerParticle_{c0}$ and the enthalpy per particle of the target structure is:
\begin{equation}
\label{deltamu}
\Delta \EnthalpyPerParticle = \EnthalpyPerParticle_{c0}-\EnthalpyPerParticle_{target}.
\end{equation}
\item Having $\Delta \EnthalpyPerParticle>0$ will establish the target as the ground state. 
However, as illustrated in Fig.~\ref{PotentialSurface_Cartoon}, $\Delta \EnthalpyPerParticle$ does not reflect the enthalpy cost to deform the target structure. 
Thus, optimizing for $\Delta \EnthalpyPerParticle$ can lead to undesirable phonon spectra. 
To overcome this problem, we incorporate quantities that enable us to modify the second derivative of the enthalpy around the target structure. 
For a fixed $\NbrPerCell$, the enthalpy per particle $\EnthalpyPerParticle(\mathbf r^\NbrPerCell; \AllLatticeVectors)$ is a function of particle positions and lattice vectors. 
The Hessian matrix of this function is calculated and its lowest non-zero eigenvalue, $\lambda_0$, is calculated. 
(In $d$ dimensions, the matrix has $d(d+1)/2$ zero-valued eigenvalues corresponding to the translation of particles and the rotation of the fundamental cell.)
Maximizing $\lambda_0$ will improve phonon stability. 
We also want to favor the smallest possible potential cut-off distance $r_c$. Therefore, we choose to maximize the objective function $\lambda_0\Delta \EnthalpyPerParticle/(1+r_c^d)$,
where $r_c^d$ is proportional to the volume of the influence sphere of the potential.
To sum up, the optimization problem is specified by the following description:
\begin{equation}
\text{maximize } \frac{\lambda_0\Delta \EnthalpyPerParticle}{1+r_c^d} \text{, subject to } \Delta \EnthalpyPerParticle>0 \text{, } \lambda_0>0 \text{, and } r_c >0.
\end{equation}
\end{enumerate}

Having defined the objective function, we use an optimizer to maximize it. 
We employ the optimizer to evaluate this objective function thousands of times using different parameters. 
Note that each objective function evaluation requires multiple inherent structure calculations.
When optimizing for this objective function, the success rate can be low. This is partially due to the fact that the objective function is neither differentiable nor continuous.
We found that the nonlinear ``Subplex'' optimization algorithm \cite{rowan1990functional} is relatively robust in optimizing this objective function.
However, we usually still need to implement the optimization hundreds of times starting from different, random sets of parameters
to ensure that we obtain the best solution in a computationally feasible way. 
To relieve the problem, we optimize for the local stability of the target structure before optimizing for the above mentioned objective function.
More precisely, we find target structure's inherent structure (which is the target structure itself if the target structure is locally stable), calculate the coordination structures of the target structure and its inherent structure, and minimize the difference between the two coordination structures.

\subsection{Verification of the Ground State}
After the optimization step, we cool, via simulated annealing, liquid configurations of particles interacting with the putative optimized potential
to absolute zero temperature to verify that the target is indeed the ground state.
To increase computational efficiency, we use relatively small systems (1 to 24 particles) in a fully deformable simulation box under periodic boundary conditions. 
We also use the thermodynamic cooling schedule, which is given by Eq.~(6) of Ref.~\onlinecite{nourani1998comparison}.

In this step, if we discover new structures that are more stable (i.e., have a lower enthalpy) than the target structure, we add them to the competitor list and return to Step~C.
If we cannot find any competitor and can find the target structure multiple times (10 times in the current implementation), 
then the target structure is deemed to be the ground state of the optimized potential.
We finally check the result by calculating the target structure's phonon spectrum and ensure that all of the phonon frequencies are real. 
When calculating the phonon spectrum, we assume that each particle has a unit mass. 
We calculate the phonon frequency squared $\omega^2$ along some trajectories between points of high symmetry in the Brillouin zone and ensure the nonnegativity condition $\omega^2 \ge 0$ for all wavevectors.
The choice of the high-symmetry points for each target structure is given in Appendix~\ref{KPoints}.

\section{Results}
\label{results}
In this section, we report optimized potentials for our target structures. 
To test the validity of each potential, we have also performed Monte Carlo or molecular-dynamics based simulated annealing on relatively large systems, as explained in detail below. 
We have also calculated the elastic constants of our target structures, which are presented in Appendix~\ref{ElasticConstants}.
The rectangular lattices and the rectangular kagome crystal are elastically anisotropic structures.
\subsection{kagome Crystal}
\begin{figure}
\begin{center}
\includegraphics[width=100mm]{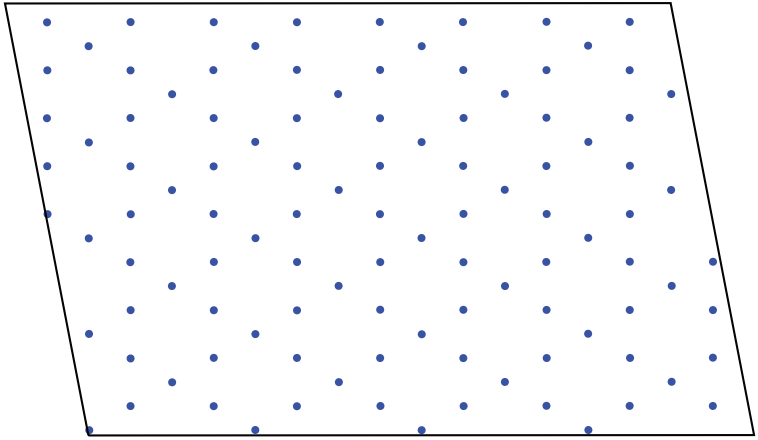}
\end{center}
\caption{(Color online) Result of a 108-particle simulated annealing for the potential given by Eq.~(\ref{Kagome_H_Eq}). This is a perfect kagome crystal.}
\label{Kagome_H_MC}
\end{figure}

\begin{figure}
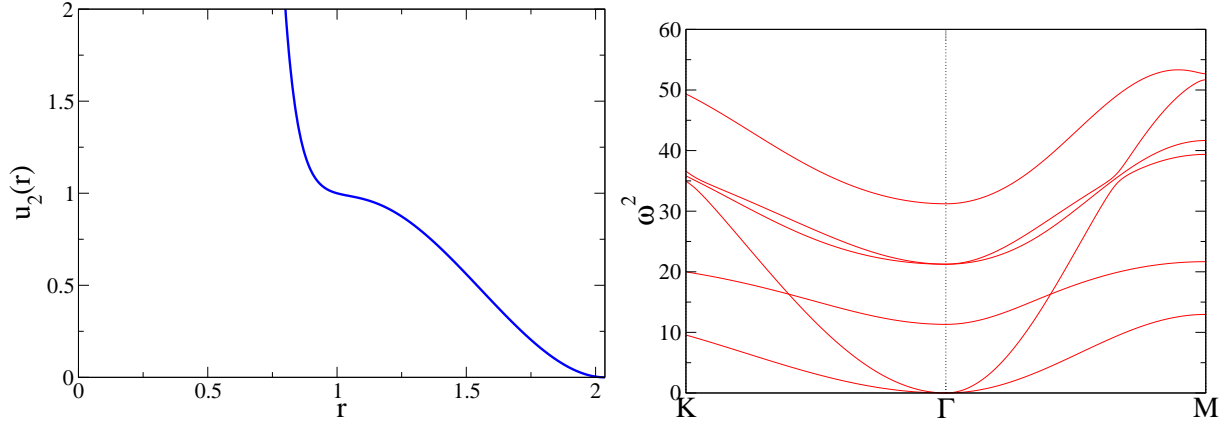

\begin{center}
\includegraphics[width=80mm]{Fig4a.eps}
\includegraphics[width=80mm]{Fig4b.eps}
\end{center}
\caption{(Color online) Left panel: The kagome potential $u_2(r)$ versus distance corresponding to Eq.~(\ref{Kagome_H_Eq}). Right panel: The phonon frequency squared $\omega^2$ vs the wavevector of the kagome crystal.}
\label{Kagome_H_result}
\end{figure}

The kagome crystal, as shown in Fig.~\ref{Kagome_H_MC}, is a 2D crystal structure obtained by removing one quater of the particles in the triangle lattice. 
The vacancies form a larger triangle lattice.
Each fundamental cell contains three particles, and each particle has four nearest neighbors. The local environments of all particles are equivalent up to rotations and translations.
At pressure $p=2.83709$, the kagome crystal is the ground state of the following potential:
\begin{equation}
u_2(r)=
\begin{cases}
\displaystyle \left(\frac{b}{r^{12}}+c_0+c_1r \right)(r-r_{c})^2,& \text{if $r<r_{c}$,} \\
0,& \text{otherwise,}
\end{cases}
\label{Kagome_H_Eq}
\end{equation}
where $b=5.9860\e{-2}$, $c_0=-1.2811$, $c_1=2.1521$, and $r_{c}=2.0364$. The potential and the phonon spectrum of the kagome crystal are shown in Fig.~\ref{Kagome_H_result}.
The ending configuration of a 108-particle simulated annealing run is shown in Fig.~\ref{Kagome_H_MC} and is seen to be the perfect kagome crystal.

\subsection{Rectangular Lattices}
Rectangular lattices are 2D Bravais lattices \cite{kittel2005introduction} in which the two lattice vectors are perpendicular but not equal in length. 
Let the lengths of two lattice vectors be $a$ and $b$; we call $b/a$ the {\it aspect ratio}. 
When $b/a \ne 1$, the rectangular lattice generally does not retain its aspect ratio when the pressure is perturbed. 
However, as shown in Appendix~\ref{rectangle_pressurerange}, for a specific class of potentials, a rectangular lattice does retain its aspect ratio in a nontrivial pressure range.

\begin{figure}
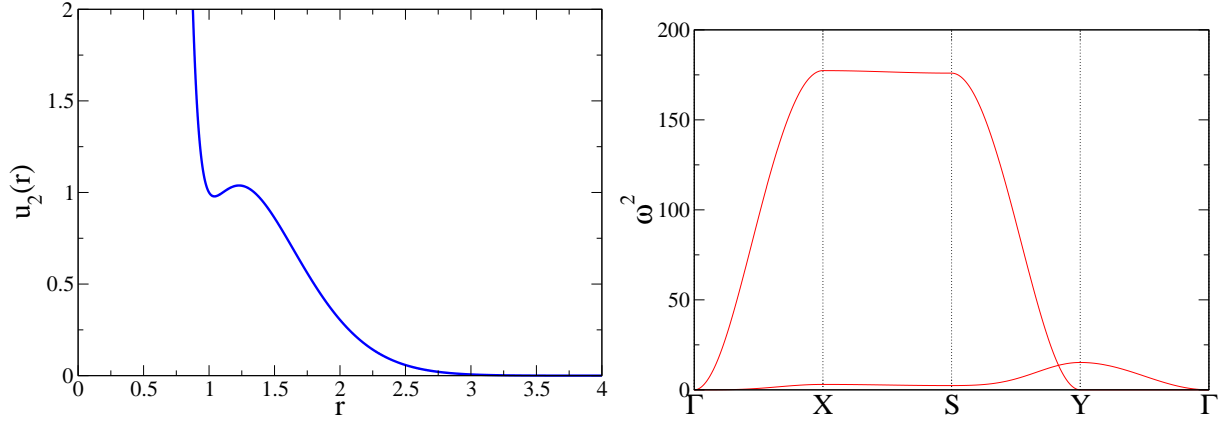

\begin{center}
\includegraphics[width=80mm]{Fig5a.eps}
\includegraphics[width=80mm]{Fig5b.eps}
\end{center}
\caption{(Color online) Left panel: Lower-order potential $u_2(r)$ vs distance for the rectangular lattice with aspect ratio $b/a=2$, corresponding to Eq.~(\ref{Rectangle_H_3_Eq}). Right panel: The phonon frequency squared $\omega^2$ vs the wavevector of the target.}
\label{Rectangle_H_3_result}
\end{figure}

\begin{figure}
\begin{center}
\includegraphics[width=100mm]{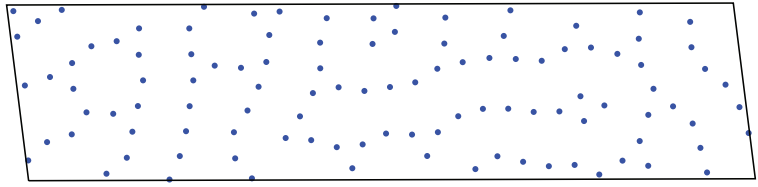}
\end{center}
\caption{(Color online) Result of a 108-particle simulated annealing for the potential given by Eq.~(\ref{Rectangle_H_3_Eq}). 
The particles show a tendency to self-assemble into the rectangular lattice with aspect ratio $b/a=2$, but many defects exist in the resulting configuration.}
\label{Rectangle_H_3_MC}
\end{figure}

We undertook to stabilize the rectangular lattice with aspect ratio $b/a=2$ using the potential form in Eq.~(\ref{function_form}). 
We found that this target structure can indeed be stabilized by the following potential at pressure $p=1.81198$:
\begin{equation}
u_2(r)=
\begin{cases}
\displaystyle \left(\frac{b}{r^{12}}+c_0+c_1r \right)\exp(-\alpha r^2)(r-r_{c})^2,& \text{if $r<r_{c}$,} \\
0,& \text{otherwise,}
\end{cases}
\label{Rectangle_H_3_Eq}
\end{equation}
where $b=2.1639\e{-2}$, $c_0=-0.26107$, $c_1=0.31488$, $\alpha=0.78857$, and $r_{c}=6.4$. 
The potential and the phonon spectrum of the rectangular lattice with aspect ratio $b/a=2$ are shown in Fig.~\ref{Rectangle_H_3_result}.
In the phonon spectrum, there is a very low branch between the $\Gamma$ and $Y$ points (defined in Appendix~\ref{KPoints}), 
indicating that there is a way to deform the target structure with very low energy cost.
The final configuration of a 108-particle simulated annealing run is shown in Fig.~\ref{Rectangle_H_3_MC}. 
Although the particles show a tendency to self-assemble to the target lattice, the ending configuration is clearly disordered, 
revealing the difficulty to crystallize particles interacting with this potential.

\begin{figure}
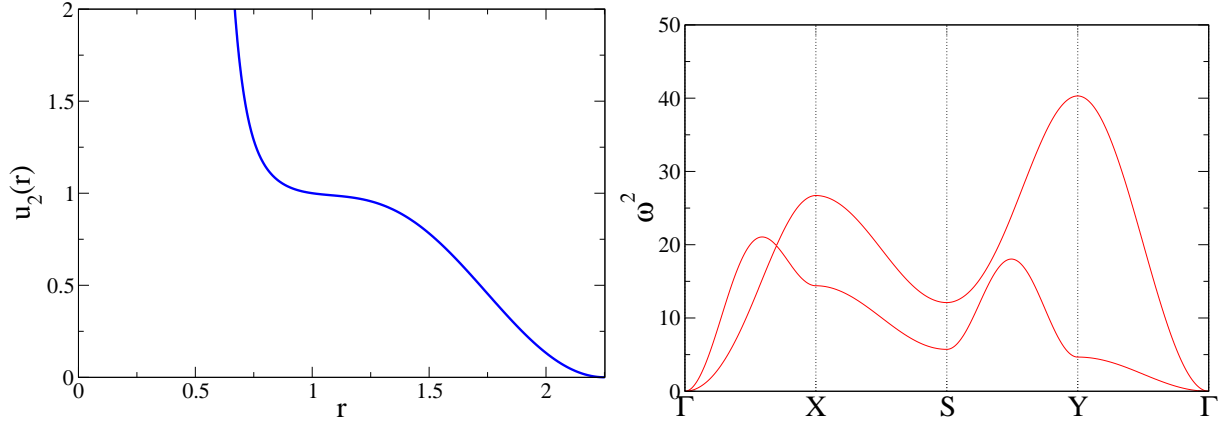

\begin{center}
\includegraphics[width=80mm]{Fig7a.eps}
\includegraphics[width=80mm]{Fig7b.eps}
\end{center}
\caption{(Color online) Left panel: Higher-order potential $u_2(r)$ versus distance for a rectangular lattice with aspect ratio $b/a=2$, corresponding to Eq.~(\ref{Rectangle_H_7_Eq}). Right panel: The phonon frequency squared $\omega^2$ vs the wavevector of the target.}
\label{Rectangle_H_7_result}
\end{figure}

\begin{figure}
\begin{center}
\includegraphics[width=100mm]{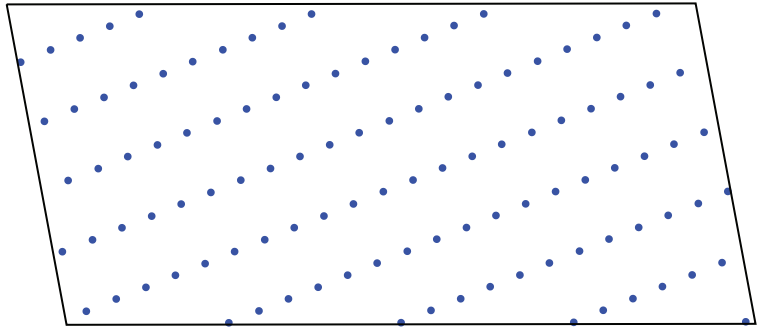}
\end{center}
\caption{(Color online) Result of a 108-particle simulated annealing for the potential given by Eq.~(\ref{Rectangle_H_7_Eq}). 
This is a perfect rectangular lattice with aspect ratio $b/a=2$.}
\label{Rectangle_H_7_MC}
\end{figure}

These results can be improved when we increase the order of the polynomial in Eq.~(\ref{function_form}). 
We found that the target can be well stabilized using the following potential at pressure $p=1.12901$:
\begin{equation}
u_2(r)=
\begin{cases}
\displaystyle \left(\frac{b}{r^{12}}+c_0+c_1r+c_2r^2+c_3r^3+c_4r^4+c_5r^5\right)\exp(-\alpha r^2)(r-r_{c})^2,& \text{if $r<r_{c}$,} \\
0,& \text{otherwise,}
\end{cases}
\label{Rectangle_H_7_Eq}
\end{equation}
where $b=3.0058\e{-3}$, $c_0=0.69293$, $c_1=-0.30361$, $c_2=9.3960\e{-2}$, $c_3=-0.36154$, $c_4=0.82231$, $c_5=4.3741\e{-2}$, $\alpha=0.44095$, and $r_{c}=2.2524$.
The potential and the phonon spectrum of the rectangular lattice with aspect ratio $b/a=2$ are shown in Fig.~\ref{Rectangle_H_7_result}.
The branch between the $\Gamma$ and $Y$ points has been lifted, suggesting that it is harder to deform the target structure.
The final configuration of a 108-particle simulated annealing run is shown in Fig.~\ref{Rectangle_H_7_MC}. 
The result is a perfect rectangular lattice with aspect ratio $b/a=2$.

\begin{figure}
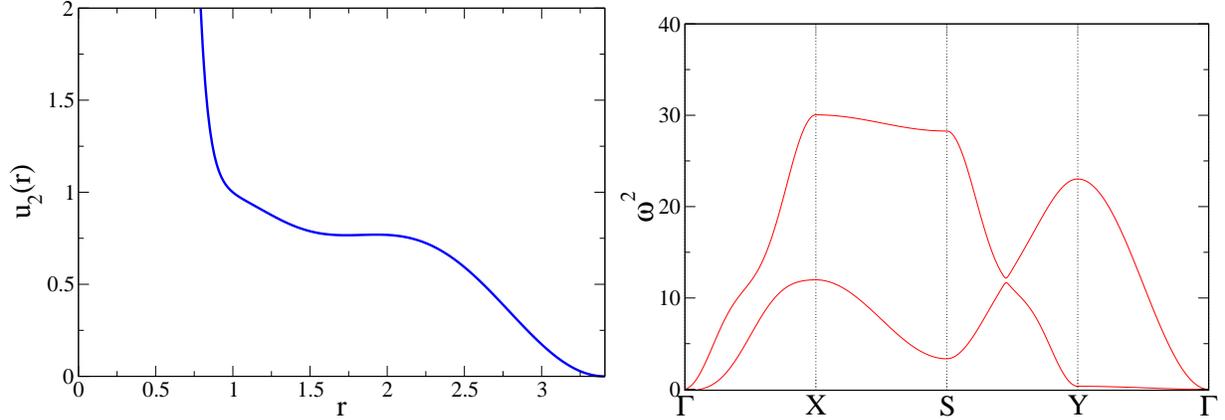

\begin{center}
\includegraphics[width=80mm]{Fig9a.eps}
\includegraphics[width=80mm]{Fig9b.eps}
\end{center}
\caption{(Color online) Left panel: The potential $u_2(r)$ vs distance for rectangular lattice with aspect ratio $b/a=\pi$, corresponding to Eq.~(\ref{Rectanglepi_H_Eq}). Right panel: The phonon frequency squared $\omega^2$ vs the wavevector of the target.}
\label{Rectanglepi_H_result}
\end{figure}

\begin{figure}
\begin{center}
\includegraphics[width=100mm]{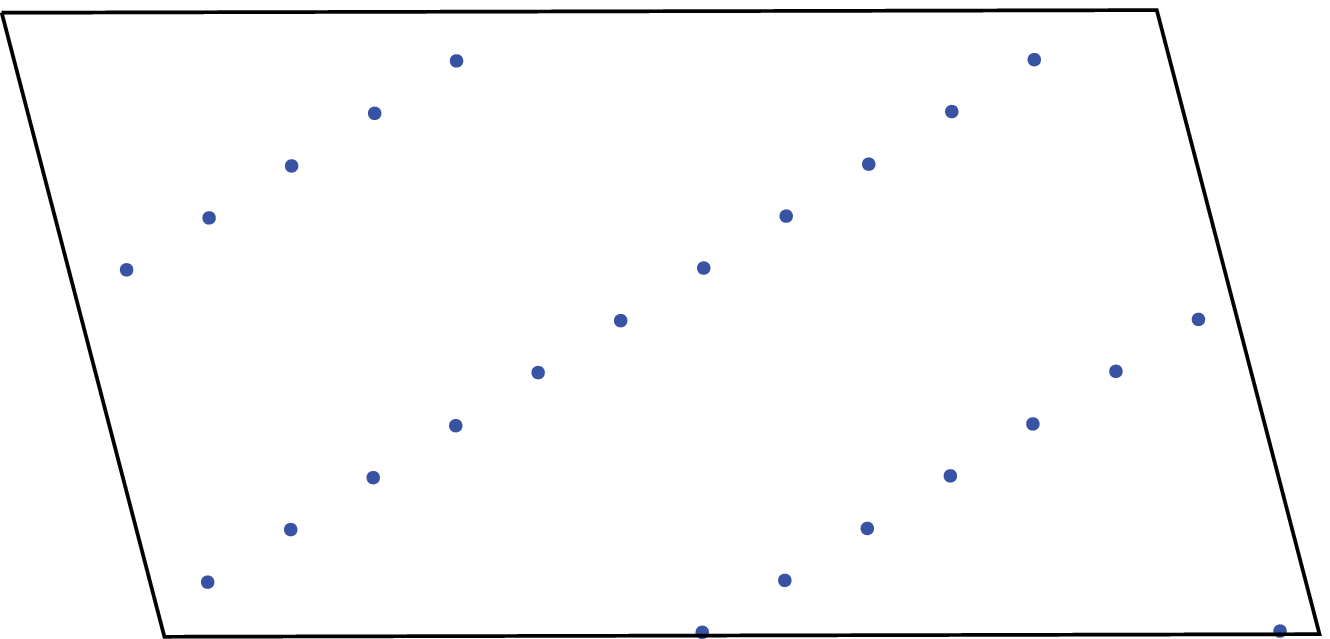}
\end{center}
\caption{(Color online) Result of a 24-particle simulated annealing for the potential given by Eq.~(\ref{Rectanglepi_H_Eq}). 
This is a perfect rectangular lattice with aspect ratio $b/a=\pi$.}
\label{Rectanglepi_H_MC}
\end{figure}

Using the optimization technique, we can also stabilize rectangular lattices with unusually large aspect ratios. 
For example, at pressure $p=1.04006$, the rectangular lattice with aspect ratio $b/a=\pi$ is the ground state of the following potential:
\begin{equation}
u_2(r)=
\begin{cases}
\displaystyle \left(\frac{b}{r^{12}}+c_0+c_1r+c_2r^2+c_3r^3+c_4r^4 \right)\exp(-\alpha r^2)(r-r_{c})^2,& \text{if $r<r_{c}$,} \\
0,& \text{otherwise,}
\end{cases}
\label{Rectanglepi_H_Eq}
\end{equation}
where $b=1.1416\e{-2}$, $c_0=-1.1117$, $c_1=3.3164$, $c_2=-3.1330$, $c_3=1.2578$, $c_4=-0.16340$, $\alpha=0.0309012$, and $r_{c}=3.4103$.
The potential and the phonon spectrum of the rectangular lattice with aspect ratio $b/a=\pi$ are shown in Fig.~\ref{Rectanglepi_H_result}.
The branch between the $\Gamma$ and $Y$ points is low, because when the aspect ratio increases, it becomes increasingly difficult to prevent the target structure from deforming.
Obtaining the target structure as a ground state using simulated annealing is also not easy. In fact, we were only able to achieve the ground state with a system of 24 particles.
The ending configuration of an 24-particle simulated annealing run is shown in Fig.~\ref{Rectanglepi_H_MC}. 
The result is a perfect rectangular lattice with aspect ratio $b/a=\pi$.

\subsection{Rectangular kagome Crystal}
\label{Result_RecKagome}

\begin{figure}
\begin{center}
\includegraphics[width=60mm]{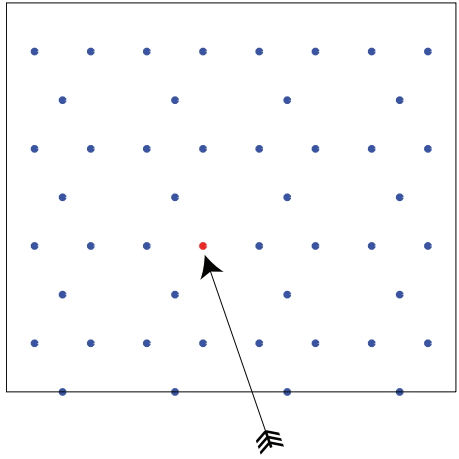}
\end{center}
\caption{(Color online) The rectangular kagome crystal structure. The particle indicated by an arrow (the red particle in colored version) has three nearest neighbors on the left and one nearest neighbor on the right, thus it is very hard to be stabilized.}
\label{RectangularKagome}
\end{figure}

\begin{figure}
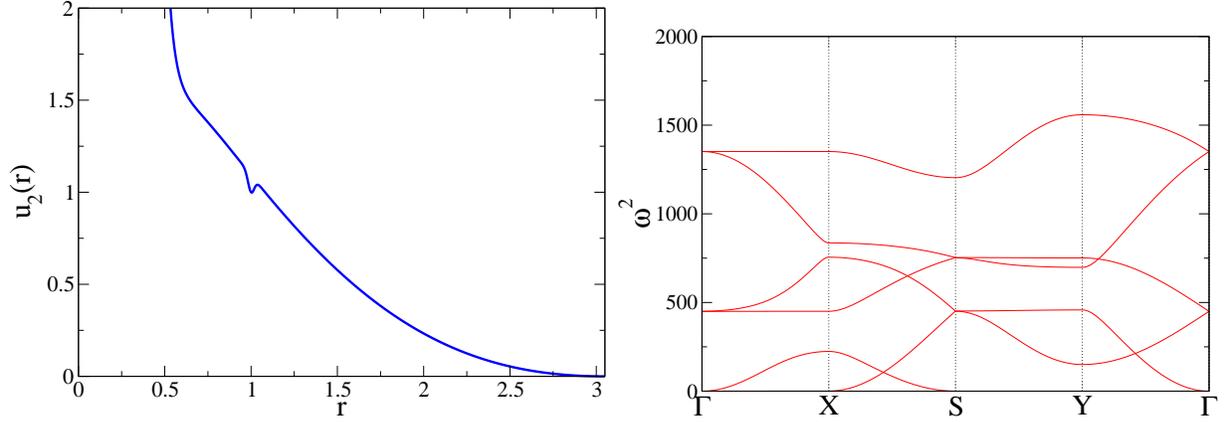

\begin{center}
\includegraphics[width=80mm]{Fig12a.eps}
\includegraphics[width=80mm]{Fig12b.eps}
\end{center}
\caption{(Color online) Left panel: The rectangular kagome potential $u_2(r)$ vs distance corresponding to Eq.~(\ref{RectangularKagome_H_Eq}). Right panel: The phonon frequency squared $\omega^2$ vs the wavevector of the rectangular kagome crystal.}
\label{RectangularKagome_H_result}
\end{figure}

\begin{figure}
\begin{center}
\includegraphics[width=100mm]{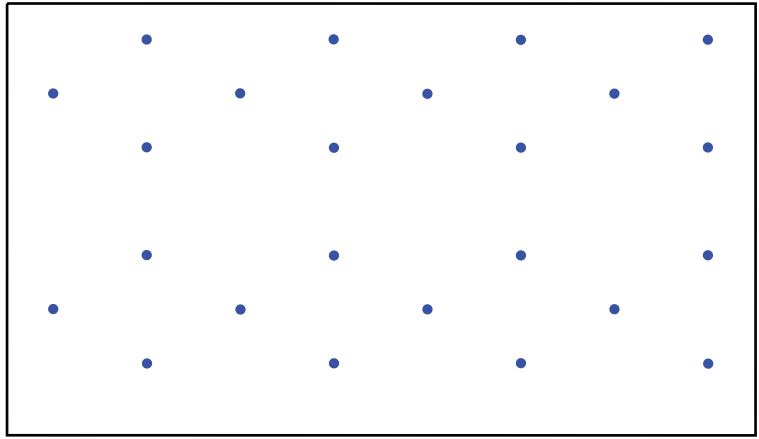}
\end{center}
\caption{(Color online) Result of a 24-particle simulated annealing for the potential given by Eq.~(\ref{RectangularKagome_H_Eq}). This is a perfect rectangular kagome crystal.}
\label{RectangularKagome_H_MC}
\end{figure}

The rectangular kagome crystal is shown in Fig.~\ref{RectangularKagome}. This crystal is similar to kagome crystal because they are both triangle lattices with vacancies, 
and each particle has four nearest neighbors. However, unlike the kagome crystal, where the vacancies are arranged in a triangle lattice, 
in the rectangular kagome crystal the vacancies are arranged in a rectangular lattice. 
Unlike all previous targets, where symmetry guarantees that the total force on each particle is zero, the local stability of some particles in the rectangular kagome crystal is not guaranteed by the symmetry.
For example, the particle indicated by an arrow in Fig.~\ref{RectangularKagome} has three nearest neighbors on the left and one nearest neighbor on the right, and thus it is not necessarily in force equilibrium. 
Accordingly, the rectangular kagome crystal is a very challenging target structure. 
In fact, we were unable to stabilize this structure using the previous potential form, which produces smooth decaying functions. 
By exploring different potential forms, we found that the rectangular kagome crystal is the ground state of the following potential at pressure $p=3.97107$:
\begin{multline}
\displaystyle u_2(r)=
\begin{cases}
\displaystyle \left(0.012352r+0.27370 \right)\exp(-0.086364 r^2)(r-3.050295)^2\\
\displaystyle +\frac{3.8032\e{-4}}{r^{12}}-\frac{1.0430\e{-2}}{r^6}\\
\displaystyle -0.092965\exp[-(\frac{r-0.99953}{0.024893})^2]+1.2956\e{-5},& \text{if $r<3.050295$,} \\
0,& \text{otherwise.}
\end{cases}
\label{RectangularKagome_H_Eq}
\end{multline}
The potential and the phonon spectrum of the rectangular kagome crystal are shown in Fig.~\ref{RectangularKagome_H_result}.
The potential contains a small Gaussian well, which is very helpful in stabilizing the particles with asymmetrical environments and forcing them to stay in the correct position.
However, this narrow well in the potential greatly increases the frequency of some phonon modes, whereas it is not helpful for other phonon modes.
Thus in the phonon spectrum, some branches are negligibly low compared to other branches.
Using this potential, we were able to get rectangular kagome crystal with simulated annealing, as shown in Fig.~\ref{RectangularKagome_H_MC}.
The presence of a small Gaussian well indicates that this isotropic pair potential is experimentally unattainable. Consequently it would be scientifically useful to determine if three-body interaction would enhance stability.


\subsection{CaF$_2$ Crystal Inhabited by a Single Particle Species}
\label{Result_CaF2}
\begin{figure}
\begin{center}
\includegraphics[width=80mm]{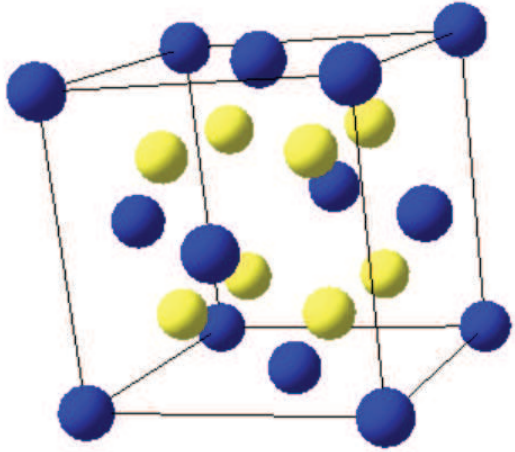}
\end{center}
\caption{(Color online) The conventional unit cell of a CaF$_2$ crystal. Blue (dark gray) spheres are Ca$^{2+}$ ions, yellow (light gray) spheres are F$^-$ ions. Particle radii are drawn proportionally to their crystal ionic radii \cite{shannon1976revised} $r$(Ca$^{2+}$)=126pm, $r$(F$^-$)=117pm.}
\label{CaF2_UnitCell}
\end{figure}

In the CaF$_2$ crystal, Ca$^{2+}$ ions are located in a face-centered cubic lattice, and F$^-$ ions fill in all the tetrahedral voids. 
A conventional unit cell of the CaF$_2$ crystal is shown in Fig.~\ref{CaF2_UnitCell}. 
Unlike previous target structures, the CaF$_2$ crystal obviously contains two kinds of particles: Each Ca$^{2+}$ ion has eight nearest neighbors while each F$^-$ ion has four nearest neighbors.
However, we found that this structure can counterintuitively be the ground state of a single-component system with the following potential at pressure $p=6.19610$:
\begin{equation}
u_2(r)=
\begin{cases}
\displaystyle \left(\frac{b}{r^{12}}+c_0+c_1r+c_2r^2+c_3r^3+c_4r^4 \right)\exp(-\alpha r^2)(r-r_{c})^2,& \text{if $r<r_{c}$,} \\
0,& \text{otherwise,}
\end{cases}
\label{CaF2_H_Eq}
\end{equation}
where $b=2.9340\e{-3}$, $c_0=0.83963$, $c_1=0.36976$, $c_2=-0.13150$, $c_3=- 2.1869\e{-3}$, $c_4=1.5010\e{-3}$, $\alpha=0.18682$, and $r_{c}=2.0564$.

\begin{figure}
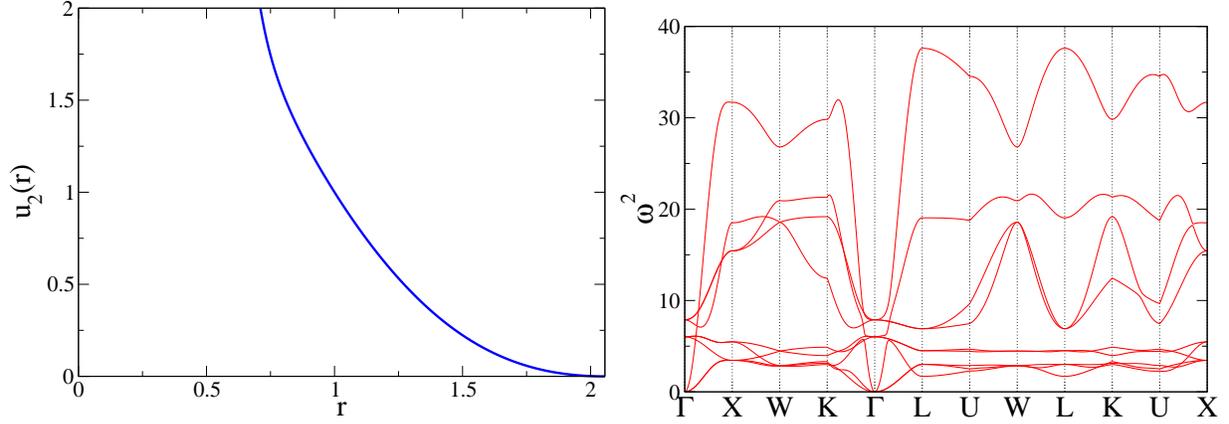

\begin{center}
\includegraphics[width=80mm]{Fig15a.eps}
\includegraphics[width=80mm]{Fig15b.eps}
\end{center}
\caption{(Color online) Left panel: The CaF$_2$ potential $u_2(r)$ vs distance corresponding to Eq.~(\ref{CaF2_H_Eq}). Right panel: The phonon frequency squared $\omega^2$ vs the wavevector of the CaF$_2$ crystal inhabited by a single particle species.}
\label{CaF2_H_result}
\end{figure}

\begin{figure}[h!]
\begin{center}
\includegraphics[width=80mm]{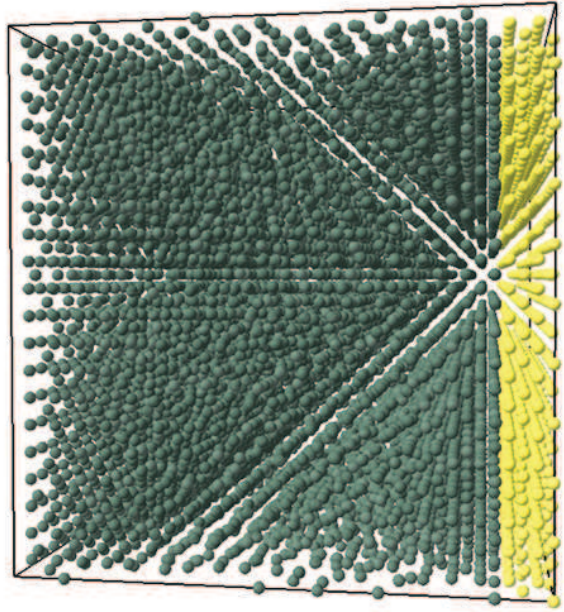}
\end{center}
\caption{Result of a 12000-particle MD based simulated annealing for the potential given by Eq.~(\ref{CaF2_H_Eq}). Yellow (light gray) particles are fixed into the CaF$_2$ structure during the simulation. Green (dark gray) particles self-assemble into the same structure.}
\label{CaF2_MD}
\end{figure}

The potential and the phonon spectrum of the target crystal are shown in Fig.~\ref{CaF2_H_result}. 
When we do simulated annealing using this potential, we rarely get the target structure when the system contains three or six particles. We were not able to achieve the ground state with larger systems. 
However, since we have tried simulated annealing using 1 to 18 particles and have never found any competitor structures with lower enthalpy, we still believe the target structure is the ground state of this potential.

To further test the validity of this potential, we have performed a molecular dynamics (MD) based simulated annealing running on graphics processing unit \cite{HOOMDWeb, anderson2008general} of 12000 particles in a fixed cubic box. The side length of the box is 10 times the side length of a CaF$_2$ conventional unit cell. Imitating the work by Rechtsman et al. \cite{rechtsman2007synthetic}, we fix 1200 particles into a layer of CaF$_2$ conventional unit cells and let the remaining 10800 particles move starting from a random sequential addition configuration with collision radius $r=0.7$. Upon slow cooling, we find that CaF$_2$ epitaxially grows from the fixed layer. The ending configuration is given in Fig.~\ref{CaF2_MD}.

\section{Conclusions and Discussion}
\label{conclusion}
To summarize, we have improved upon previous inverse statistical mechanical optimization techniques. 
By finding the inherent structures of each competitor structure, we are able to define an improved objective function for optimization, 
thus overcoming difficulties involved in previous energy difference optimizations or pressure range optimizations. 
With this optimization technique, we have designed isotropic pair potentials so that the kagome crystal, rectangular lattices with aspect ratios of 2 and $\pi$, 
the rectangular kagome crystal, and the structure of the CaF$_2$ crystal inhabited by a single particle species become the unique ground states.

By finding potentials that can stabilize these target structures as unique ground states, we have demonstrated the robustness of our method.
Our potential which stabilizes the kagome crystal showcases our improvement over previous inverse work \cite{Sloane2003Theta}
by being comparably simple to a potential found using the forward approach \cite{batten2011novel}.
Moreover, by being able to design isotropic pair potentials for the rectangular kagome crystal and CaF$_2$ crystal inhabited by a single particle species, 
we have also demonstrated that the new method can handle target structures that contain particles in different or asymmetrical local environments.

The rectangular lattices are very simple examples of a much broader family of crystal structures with lower elastic symmetry. 
Their low elastic symmetries make them spatially scale differently in different directions when the pressure changes. 
To our knowledge, none of the target structures in this family have been stabilized with pressure range optimization. 
One structure in this family, the 3D simple hexagonal lattice, has been stabilized previously \cite{rechtsman2006self} using energy difference optimization. 
However, since the result of energy difference optimization is sensitive to structurally close competitors (e.g. other rectangular lattices with slightly different aspect ratios), 
one cannot precisely control the aspect ratio. 
In contrast, our new method allows us to precisely specify large unusual aspect ratios (for example, $\pi$) when targeting this family of structures.

All of our target structures are stabilized as unique ground states.
In the application of inverse statistical mechanics to spin systems \cite{distasio2013designer}, the possible outcomes for a given target configuration
were organized into the following three solution classes: unique (nondegenerate) ground state (class I), degenerate ground states with the same two-point correlation functions (class II), 
and solutions not contained in the previous two classes (class III). All of the target structures considered in this paper fall within class I.
A simple thought experiment yields an example of class III solutions. 
Since the fcc and hcp crystals have different coordination structures, they cannot fall within class II.
If we limit the range of the pair potential to be between the nearest and next nearest neighbors, we will not be able to distinguish these target pairs from one another and thus they cannot belong to class I.
Therefore they will fall within class III.
It would be interesting to see if one can stabilize any class II solutions for many particle system.

If a target crystal structure falls within class I, then what are the necessary functional characteristics of the potential? 
For example, can we stabilize a particular target with monotonic pair potential? What is the minimum range (cut-off) of the pair potential? 
While rigorous answers to these questions are beyond the scope of the present paper, we can offer some general principles that may provide guidance 
in determining whether certain target structures can be achieved as ground states by a particular class of radial pair potentials.
Let us consider the first question.
Our experience is that, there are target structures where the symmetry does not guarantee that the total force on each particle is zero (for example, rectangular kagome crystal). 
Target structures of this kind cannot be stabilized with monotonic radial pair potentials. 
Other target structures can be stabilized with either monotonic potentials or potentials with wells.
For example, the diamond crystal has been stabilized with both a monotonic potential \cite{marcotte2012designed} and a potential with wells \cite{rechtsman2007synthetic}.

Concerning the second question,
the minimum range of the pair potential varies for different targets, but is usually comparable to the longest diagonal length of the fundamental cell $l_{dia}$ of the target crystal. 
It seems that in order for the particles to self-assemble into the target crystal, 
the pair potential only needs to encode coordination information within a range comparable in size to the fundamental cell 
(since the crystal is the replication of the fundamental cell under periodic boundary conditions).
For certain relatively symmetric target structures, the fundamental cell consists of particle subsets that differ only by translations, rotations, and inversions.
Thus, the pair potentials for these targets may only require a cut-off distance $r_c$ that is shorter than the longest diagonal of the fundamental cell.
Examples include our kagome and CaF$_2$ potentials, a previously designed body-centered-cubic potential \cite{rechtsman2006self}, 
and a kagome potential found with the forward approach \cite{batten2011novel}.
For certain relatively challenging target such as the rectangular kagome crystal (it is challenging because of the reasons explained in Sec.~\ref{Result_RecKagome}), 
the range of the potential can be somewhat longer than the longest diagonal of the fundamental cell.
In fact, the length and symmetry of the fundamental cell are the most important factors determining the required range of the potential.
This is demonstrated by the CaF$_2$ crystal inhabited by a single particle species, which is symmetric but challenging (because it contains particles in different local environments).
The optimized potential that we have obtained here contains a relatively high-order polynomial, but its range is surprisingly short.
Table~\ref{cut-off2} summarizes the minimal cut-off distance $r_c$ that we found for the targets considered in this paper.
To further support the notion that the minimal potential cut-off distance $r_c$ need only be comparable in size to the longest diagonal of the fundamental cell, we have also generated short-ranged isotropic pair potentials using our algorithm for several other simpler targets. 
Except for the fcc crystal, all of them have been stabilized before,
including the 2D honeycomb crystal \cite{rechtsman2005optimized, rechtsman2006designed, marcotte2011optimized, marcotte2011unusual}, 
2D square lattice \cite{rechtsman2006designed, marcotte2011optimized, marcotte2011unusual}, 
3D bcc lattice \cite{rechtsman2006self}, 
3D simple cubic lattice \cite{rechtsman2006self},
3D diamond crystal \cite{rechtsman2007synthetic, marcotte2012designed},
and 3D fcc lattice. 
We see in Table~\ref{cut-off1} that the potential cut-off distances are indeed comparable in size to the longest diagonal of the fundamental cell, 
which is consistent with our results for the more complicated targets listed in Table~\ref{cut-off2}.

\begin{table}[h]
\setlength{\tabcolsep}{12pt}
\caption{Isotropic pair potential cut-off $r_{c}$, longest diagonal length of the fundamental cell $l_{dia}$, and their ratio of targets reported in Sec.~\ref{results}. The nearest neighbor distance is 1.}
\begin{tabular}{c c c c}
\hline
Target Structure & $r_{c}$ & $l_{dia}$ & $r_{c}/l_{dia}$ \\ \hline
kagome & 2.04 & $2\sqrt{3}$ & 0.59\\
Rectangular lattice $b/a=2$ & 2.25 & $\sqrt{5}$ & 1.00\\
Rectangular lattice $b/a=\pi$ & 3.41 & $\sqrt{\pi^2+1}$ & 1.03\\
CaF$_2$ single species & 2.06 & 4 & 0.52\\
Rectangular kagome & 3.05 & $\sqrt{7}$ & 1.15\\
\hline
\end{tabular}
\label{cut-off2}
\end{table}

\begin{table}
\setlength{\tabcolsep}{12pt}
\caption{Application of our current optimization scheme to stabilize simpler targets with potentials having a minimal cut-off distance $r_c$ for the family of potential functions indicated in Eq.~(\ref{function_form}). Except for the fcc lattice, all of the targets have been stabilized before \cite{rechtsman2005optimized, rechtsman2006designed, marcotte2011optimized, marcotte2011unusual, rechtsman2006self, rechtsman2007synthetic, marcotte2012designed}.
Isotropic pair potential cut-off $r_{c}$, longest diagonal length of the fundamental cell $l_{dia}$, and their ratio are listed. The nearest neighbor distance is 1.}
\begin{tabular}{c c c c}
\hline
Target Structure & $r_{c}$ & $l_{dia}$ & $r_{c}/l_{dia}$ \\ \hline
Honeycomb & 2.53 & 3 & 0.84\\
Square & 1.87 & $\sqrt{2}$ & 1.32 \\
bcc & 1.24 & $\sqrt{11/3}$ & 0.65 \\
Simple Cubic & 1.54 & $\sqrt{3}$ & 0.89 \\
Diamond & 2.46 & $4$ & 0.62 \\
fcc & 1.77 & $\sqrt{6}$ & 0.72 \\
\hline
\end{tabular}
\label{cut-off1}
\end{table}

What are the limitations of isotropic pair potentials in achieving targeted ground states? 
In other words, given a target structure, how can we tell whether an isotropic pair potential can stabilize it or not? 
Since the enthalpy per particle is determined by the coordination numbers $Z_j$ and specific volume $\VolumePerParticle$ in Eq.~(\ref{enthalpyperparticle}), 
a target structure cannot be stabilized by isotropic pair potentials as unique ground state if its coordination numbers and specific volume are identical to that of another structure, 
or are a weighted average of other structures \cite{cohn2009algorithmic}.
This implies, for example, that chiral targets with only one type of handedness cannot be uniquely stabilized by isotropic pair interactions \cite{torquato2009inverse}.
Besides the target structures that are disproved by this theorem, are all other structures realizable by isotropic pair interactions?
Seeking a full answer to this question will be a direction of future research.

The entire set of possible target structures extends far beyond what has been examined. 
Specifically, this includes challenging structures such as ``tunneled'' crystals \cite{torquato2007toward} characterized by a high concentration of chains of vacancies
as well as the graphite crystal, to mention a few examples. 
A direction for future research is to either stabilize them with the simplest possible radial potentials or to prove that they cannot be stabilized with such interactions, 
which may require us to improve the current algorithm.
We are also interested in expanding our method to stabilize multicomponent systems and systems containing particles with anisotropic interactions \cite{torquato2009inverse}. 

\begin{acknowledgments}
We are very grateful to {\'E}tienne Marcotte for many helpful discussions and Steven Atkinson for his careful reading of the manuscript.
This research was supported by the U.S. Department of Energy, Office of Basic Energy Sciences, Division of Materials Sciences and Engineering under Award No. DE-FG02-04-ER46108.
\end{acknowledgments}

\appendix
\section{Crystal Structure and Theta Series of Target Structures}
In this appendix, we provide the vectors that specify the target crystal structure as well as the corresponding partial theta series defined generally by Eq.~(\ref{ThetaSeries}).
\label{targets_structure}
\subsection{kagome Crystal}
The kagome crystal is a 2D crystal whose fundamental lattice vectors can be specified as follows:

\begin{equation}
\mathbf a_1=2 \mathbf i \text{ and }
\mathbf a_2=\mathbf i + \sqrt{3} \mathbf j.
\end{equation}
Its reciprocal lattice vectors are
\begin{equation}
\mathbf b_1=\pi \mathbf i - \frac{\pi}{\sqrt{3}} \mathbf j \text{ and }
\mathbf b_2=\frac{2\pi}{\sqrt{3}} \mathbf j.
\end{equation}
Each fundamental cell contains 3 particles, located at the positions
\begin{equation}
\begin{matrix*}[l]
\displaystyle \mathbf r_1=\frac{1}{2} \mathbf a_1 = \mathbf i \text{, } \\[0.5em]
\displaystyle \mathbf r_2=\frac{1}{2} \mathbf a_2 = \frac{1}{2}\mathbf i + \frac{\sqrt{3}}{2} \mathbf j \text{, and } \\[0.5em]
\displaystyle \mathbf r_3=\frac{1}{2} \mathbf a_1+\frac{1}{2} \mathbf a_2=\frac{3}{2}\mathbf i + \frac{\sqrt{3}}{2} \mathbf j.
\end{matrix*}
\end{equation}

The first few terms of its theta series are
\begin{equation}
\theta(q)=1+4q+4q^3+6q^4+8q^7+4q^9 + \cdots .
\end{equation}

\subsection{Rectangular Lattice with aspect ratio $t$}
Rectangular lattices are 2D crystals whose fundamental lattice vectors can be specified as follows:
\begin{equation}
\mathbf a_1=\mathbf i \text{ and }
\mathbf a_2=t \mathbf j.
\end{equation}
Its reciprocal lattice vectors are
\begin{equation}
\mathbf b_1=2\pi \mathbf i\text{ and }
\mathbf b_2=\frac{2\pi}{t} \mathbf j.
\end{equation}
Each fundamental cell contains 1 particle, located at the positions
\begin{equation}
\mathbf r_1=\mathbf 0
\end{equation}

The first few terms of its theta series are
\begin{equation}
\theta(q)=1+2q+2q^4+2q^9+ \cdots \\
+2q^{t^2}+4q^{t^2+1}+4q^{t^2+4} + \cdots .
\end{equation}

\subsection{Rectangular kagome Crystal}
The rectangular kagome crystal is a 2D crystal whose fundamental lattice vectors can be specified as follows:
\begin{equation}
\mathbf a_1=2 \mathbf i \text{ and }
\mathbf a_2=\sqrt{3} \mathbf j.
\end{equation}
Its reciprocal lattice vectors are
\begin{equation}
\mathbf b_1=\pi \mathbf i\text{ and }
\mathbf b_2=\frac{2\pi}{\sqrt{3}}\mathbf j.
\end{equation}
Each fundamental cell contains 3 particles, located at the positions
\begin{equation}
\begin{matrix*}[l]
\displaystyle \mathbf r_1=\frac{1}{2} \mathbf a_1 = \mathbf i \text{, } \\[0.5em]
\displaystyle \mathbf r_2=\frac{1}{4} \mathbf a_1 + \frac{1}{2} \mathbf a_2= \frac{1}{2}\mathbf i + \frac{\sqrt{3}}{2} \mathbf j \text{, and } \\[0.5em]
\displaystyle \mathbf r_3=\frac{3}{4} \mathbf a_1 + \frac{1}{2} \mathbf a_2= \frac{3}{2}\mathbf i + \frac{\sqrt{3}}{2} \mathbf j.
\end{matrix*}
\end{equation}

The first few terms of its theta series are
\begin{equation}
\theta(q)=1+4q+\frac{14}{3}q^3+\frac{14}{3}q^4+\frac{28}{3}q^7+4q^9 + \cdots .
\end{equation}

\subsection{CaF$_2$ Crystal Inhabited by a Single Particle Species}
The CaF$_2$ crystal inhabited by a single particle species is a 3D crystal whose fundamental lattice vectors can be specified as follows:
\begin{equation}
\mathbf a_1=\frac{2}{\sqrt{3}}( \mathbf i + \mathbf j) \text{, }
\mathbf a_2=\frac{2}{\sqrt{3}}( \mathbf i + \mathbf k) \text{, and }
\mathbf a_3=\frac{2}{\sqrt{3}}( \mathbf j + \mathbf k).
\end{equation}
Its reciprocal lattice vectors are
\begin{equation}
\mathbf b_1=\sqrt{\frac{3}{4}}\pi (\mathbf i + \mathbf j - \mathbf k)\text{, }
\mathbf b_2=\sqrt{\frac{3}{4}}\pi (-\mathbf i + \mathbf j + \mathbf k)\text{, and }
\mathbf b_3=\sqrt{\frac{3}{4}}\pi (\mathbf i - \mathbf j + \mathbf k).
\end{equation}
Each fundamental cell contains 3 particles, located at the positions
\begin{equation}
\begin{matrix*}[l]
\displaystyle \mathbf r_1=\mathbf 0 \text{, } \\[0.5em]
\displaystyle \mathbf r_2=\frac{\mathbf a_1 +\mathbf a_2 +\mathbf a_3}{4} = \frac{\mathbf i+\mathbf j+\mathbf k}{\sqrt{3}} \text{, and } \\[0.5em]
\displaystyle \mathbf r_3=\frac{3(\mathbf a_1 +\mathbf a_2 +\mathbf a_3)}{4}=\sqrt{3}(\mathbf i+\mathbf j+\mathbf k).
\end{matrix*}
\end{equation}

The first several terms of its theta series are
\begin{equation}
\theta(q)=1+\frac{16}{3}q+4q^{4/3}+12q^{8/3}+16q^{11/3}+\frac{16}{3}q^4+6q^{16/3}+16q^{19/3}+16q^{20/3}+24q^8+\frac{64}{3}q^9 + \cdots .
\end{equation}

\section{Definition of High-Symmetry Points in the Brillouin Zone}
\label{KPoints}
When ascertaining the phonon spectrum of a crystal, we calculate the phonon frequency squared $\omega ^2$ along certain trajectories between points of high symmetry in the Brillouin zone. For different crystals, the points of high symmetry are described below.
\subsection{2D kagome Crystal}
The points of high symmetry of 2D kagome crystal are
\begin{equation}
K = \frac{1}{2} \mathbf b_1 \text{, }
\Gamma = \mathbf 0 \text{, and }
M = \frac{1}{3} (\mathbf b_1+ \mathbf b_2),
\end{equation}
where $\mathbf b_1$ and $\mathbf b_2$ are reciprocal lattice vectors.
\subsection{2D Rectangular Lattices and Rectangular kagome Crystal}
The points of high symmetry of 2D rectangular lattices and rectangular kagome crystal are
\begin{equation}
\Gamma = \mathbf 0 \text{, }
X = \frac{1}{2} \mathbf b_1 \text{, }
S = \frac{1}{2} (\mathbf b_1+ \mathbf b_2) \text{, and }
Y = \frac{1}{2} \mathbf b_2,
\end{equation}
where $\mathbf b_1$ and $\mathbf b_2$ are reciprocal lattice vectors.
\subsection{CaF$_2$ Crystal Inhabited by a Single Particle Species}
The points of high symmetry of CaF$_2$ crystal inhabited by a single particle species are
\begin{equation}
\begin{matrix*}[l]
\displaystyle \Gamma = \mathbf 0 \text{, } \\[0.5em]
\displaystyle X = \frac{1}{2} (\mathbf b_1+\mathbf b_3) \text{, } \\[0.5em]
\displaystyle W = \frac{1}{4} (2\mathbf b_1+\mathbf b_2+3\mathbf b_3) \text{, } \\[0.5em]
\displaystyle K = \frac{3}{8} (\mathbf b_1+\mathbf b_2+2\mathbf b_3) \text{, } \\[0.5em]
\displaystyle L = \frac{1}{2} (\mathbf b_1+\mathbf b_2+\mathbf b_3) \text{, and } \\[0.5em]
\displaystyle U = \frac{1}{8} (5\mathbf b_1+4\mathbf b_2+5\mathbf b_3) \text{, }
\end{matrix*}
\end{equation}
where $\mathbf b_1$, $\mathbf b_2$, and $\mathbf b_3$ are reciprocal lattice vectors.

\section{Definition of the ``Difference'' Between Two Coordination Structures}
\label{coordination_difference}
The coordination structure of a crystal is characterized by coordination numbers $Z_j$ for different distances $r_j$, as defined in Sec.~\ref{technique}.
The coordination numbers and distances of a crystal structure can be summarized into an infinite table, which consists of infinite number of ``rows''. 
Each row contains a distance $r$ and the average number of neighbors $Z$ at that distance.
We have defined a ``difference'' between two coordination structures. To calculate it, we use the following steps:
\begin{enumerate}
\item Rows of the two coordination structures, $\{r, Z\}$, are combined into pairs by the following rules:
\begin{enumerate}
\item The first unpaired rows of the two coordination structures are paired if their coordination numbers are equal. 
\item If their coordination numbers are not equal, let the row with larger coordination number be $\{r_{large}, Z_{large}\}$ and the row with smaller coordination number be $\{r_{small}, Z_{small}\}$. 
The row with the larger coordination number, $\{r_{large}, Z_{large}\}$, is split into two rows: A row $\{r_{large}, Z_{small}\}$ and another row $\{r_{large}, Z_{large}-Z_{small}\}$.
The former row is paired with $\{r_{small}, Z_{small}\}$. The latter row will be paired later.
\item Return to step (a) unless enough pairs are obtained.
\end{enumerate}
For example, to combine the coordination structure of rectangular kagome crystal and that of kagome crystal into pairs of rows, we do the following. 
To illustrate the process, let us denote a row from the rectangular kagome crystal as $\{r, Z\}_r$, and a row from the kagome crystal as $\{r, Z\}_k$
\begin{enumerate}
\item The first row of the coordination structure of rectangular kagome crystal, $\{1, 4\}_r$, is paired with the first row of the coordination structure of kagome crystal, $\{1, 4\}_k$.
\item The second row of the coordination structure of rectangular kagome crystal, $\{\sqrt{3}, 14/3\}_r$, is split into two rows: a row $\{\sqrt{3}, 4\}_r$ will be paired with the second row from the kagome crystal ($\{\sqrt{3}, 4\}_k$), the other row $\{\sqrt{3}, 2/3\}_r$ will be paired later.
\item The next row from the kagome crystal, $\{2, 6\}_k$, is split into two rows: a row $\{2, 2/3\}_k$ to be paired with the remaining row from the rectangular kagome crystal, $\{\sqrt{3}, 2/3\}_r$, and another row $\{2, 16/3\}_k$ to be paired later.
\item The remaining row from the kagome crystal, $\{2, 16/3\}_k$, is split into two rows: $\{2, 14/3\}_k$ and $\{2, 2/3\}_k$. The former is paired with the third row from the rectangular kagome crystal, $\{2, 14/3\}_r$. The latter remains to be paired.
\item continue this process until enough pairs are obtained. The first several obtained pairs are:
\begin{center}
\begin{table}
\caption{The first several pairs of the coordination structure (radial distance and associated coordination number) for the kagome and rectangular kagome crystals.}
\setlength{\tabcolsep}{12pt}
\begin{tabular}{c c}
\hline
Rectangular Kagome Crystal & Kagome Crystal\\
\hline
$\{1, 4\}_r$ & $\{1, 4\}_k$\\
$\{\sqrt{3}, 4\}_r$ & $\{\sqrt{3}, 4\}_k$\\
$\{\sqrt{3}, 2/3\}_r$ & $\{2, 2/3\}_k$\\
$\{2, 14/3\}_r$ & $\{2, 14/3\}_k$\\
$\{\sqrt{7}, 2/3\}_r$ & $\{2, 2/3\}_k$\\
$\{\sqrt{7}, 8\}_r$ & $\{\sqrt{7}, 8\}_k$\\
$\{\sqrt{7}, 2/3\}_r$ & $\{3, 2/3\}_k$\\
\multicolumn{2}{c}{......} \\
\hline
\end{tabular}
\end{table}
\end{center}
\end{enumerate}
\item The distance between two coordination structures is given by:
\begin{equation}
D=\sum_{\text{all pairs $\{r_a, Z_a\}$ and $\{r_b, Z_b\}$}} Z_a(r_a-r_b)^2 \exp(-r_a)
\end{equation}
In our implementation, the summation is truncated at $r=5$.
\end{enumerate}

This definition of distance $D$ has the following properties:
\begin{enumerate}
\item $D \ge 0$. $D=0$ if and only if the two coordination structures are identical.
\item A infinitesimally distorted structure of an original structure has a coordination structure which has an infinitesimal distance to the coordination structure of the original structure.
\end{enumerate}

\section{Elastic Properties of Target Structures}
\label{ElasticConstants}

We have also calculated the elastic constants of our target structures. To illustrate the concept of elastic constants, consider a small, affine deformation of the target structure:
\begin{equation}
\mathbf x = (\mathbf I + \mathbf \epsilon) \mathbf x_0,
\end{equation}
where $\mathbf x_0$ is the original location, $\mathbf x$ is the new location, $\mathbf I$ is a unit second-order tensor, and $\mathbf \epsilon$ is a small second-order tensor, called ``strain tensor''.
The elastic constants $C_{ijkl}$ are defined as:
\begin{equation}
C_{ijkl}= \frac{\partial^2 H}{ \partial \epsilon_{ij} \partial \epsilon_{kl}}.
\end{equation}

The elastic constants of our target structures are presented below.

\subsection{2D Isotropic Target}
The kagome crystal is a 2D isotropic crystal. Its elastic constants are determined by two independent constants, e.g., its Young's modulus $E$ and Poisson's ratio $\nu$:
\begin{equation}
\begin{pmatrix}
C_{1111} & C_{1122} & C_{1112} \\
C_{2211} & C_{2222} & C_{2212} \\
C_{1211} & C_{1222} & C_{1212}
\end{pmatrix}
=
\frac{E}{1-\nu^2}
\begin{pmatrix}
1 & \nu & 0 \\
\nu & 1 & 0 \\
0 & 0 & \frac{1-\nu}{2}
\end{pmatrix}
.
\end{equation}

With the pair potential in Eq.~(\ref{Kagome_H_Eq}), under pressure $p=2.83709$, the kagome crystal has elastic constants $E=23.61$ and $\nu=0.4594$.

\subsection{2D Orthotropic Targets}
The rectangular lattices and the rectangular kagome crystal are 2D orthotropic crystals. Their elastic constants are determined by four independent constants, $E_x$, $E_y$, $G$, and $\nu_{xy}$:
\begin{equation}
\begin{pmatrix}
C_{1111} & C_{1122} & C_{1112} \\
C_{2211} & C_{2222} & C_{2212} \\
C_{1211} & C_{1222} & C_{1212}
\end{pmatrix}
=
\frac{1}{1-\nu_{xy}\nu_{yx}}
\begin{pmatrix}
E_x & \nu_{yx}E_x & 0 \\
\nu_{xy}E_y & E_y & 0 \\
0 & 0 & G(1-\nu_{xy}\nu_{yx})
\end{pmatrix}
,
\end{equation}
where $\nu_{yx}=\nu_{xy}E_y/E_x$.

With the pair potential in Eq.~(\ref{Rectangle_H_3_Eq}), under pressure $p=1.81198$, the rectangular lattice with aspect ratio 2 has elastic constants $E_x=27.31$, $E_y=7.17$, $G=0.01$, and $\nu_{xy}=0.4751$.

With the pair potential in Eq.~(\ref{Rectangle_H_7_Eq}), under pressure $p=1.12901$, the rectangular lattice with aspect ratio 2 has elastic constants $E_x=7.19$, $E_y=17.60$, $G=2.33$, and $\nu_{xy}=0.2277$.

With the pair potential in Eq.~(\ref{Rectanglepi_H_Eq}), under pressure $p=1.04006$, the rectangular lattice with aspect ratio $\pi$ has elastic constants $E_x=3.98$, $E_y=16.91$, $G=0.27$, and $\nu_{xy}=0.1296$.

With the pair potential in Eq.~(\ref{RectangularKagome_H_Eq}), under pressure $p=3.97107$, the rectangular kagome crystal has elastic constants $E_x= 177.9$, $E_y=177.5$, $G=65.3$, and $\nu_{xy}=0.3596$.

\subsection{3D Isotropic Target}
The CaF$_2$ crystal inhabited by a single particle species is a 3D cubic crystal. Its elastic constants are determined by three independent constants, $E$, $\nu$, and $A$:
\begin{multline}
\begin{pmatrix}
C_{1111} & C_{1122} & C_{1133} & C_{1123} & C_{1131} & C_{1112} \\
C_{2211} & C_{2222} & C_{2233} & C_{2223} & C_{2231} & C_{2212} \\
C_{3311} & C_{3322} & C_{3333} & C_{3323} & C_{3331} & C_{3312} \\
C_{2311} & C_{2322} & C_{2333} & C_{2323} & C_{2331} & C_{2312} \\
C_{3111} & C_{3122} & C_{3133} & C_{3123} & C_{3131} & C_{3112} \\
C_{1211} & C_{1222} & C_{1233} & C_{1223} & C_{1231} & C_{1212} 
\end{pmatrix}
= \\
\frac{E}{(1+\nu)(1-2\nu)}
\begin{pmatrix}
1-\nu & \nu & \nu & 0 & 0 & 0\\
\nu & 1-\nu & \nu & 0 & 0 & 0 \\
\nu & \nu & 1-\nu & 0 & 0 & 0 \\
0 & 0 & 0 & A(1-2\nu)/2 & 0 & 0 \\
0 & 0 & 0 & 0 & A(1-2\nu)/2 & 0 \\
0 & 0 & 0 & 0 & 0 & A(1-2\nu)/2
\end{pmatrix}
.
\end{multline}

With the pair potential in Eq.~(\ref{CaF2_H_Eq}), under pressure $p=6.19610$, the CaF$_2$ crystal inhabited by a single particle species has elastic constants $E=2.1835$, $\nu=0.4753$, and $A=2.51$.

\section{Stabilizing a Rectangular Lattice Over a Pressure Range}
\label{rectangle_pressurerange}
The rectangular lattices do not naturally have a stable pressure range because of their anisotropic elastic property. 
When the pressure changes, the two sides of the rectangular unit cell may change disproportionally, thus the aspect ratio may also change and the structure changes according to our definition.
However, we can make ``corrections'' to the potential to make sure that the aspect ratio does not change over a pressure range. Imagine a rectangular lattice with one side length $a$ 
and the other side length $b=at$, thus the aspect ratio is $t$. In order for the rectangular lattice with aspect ratio $t$ to be stable in a pressure range, when pressure $p$ changes in the range, 
$a$ or $b$ can change while the aspect ratio $t$ must not change. The enthalpy of the target is given by:
\begin{equation}
\label{correction_1}
H= \sum_{(i, j) \neq (0, 0)} u_2\left (\sqrt{i^2+(jt)^2}a\right)+ pa^2t.
\end{equation}
When the structure is stable, the partial derivatives of enthalpy are zero. Thus:
\begin{equation}
\label{correction_2}
\frac{\partial H}{\partial a}=\sum_{(i, j) \neq (0, 0)} u_2'\left (\sqrt{i^2+(jt)^2}a\right)\sqrt{i^2+(jt)^2}+2pat=0,
\end{equation}
and:
\begin{equation}
\label{correction_3}
\frac{\partial H}{\partial t}=\sum_{(i, j) \neq (0, 0)} u_2'\left (\sqrt{i^2+(jt)^2}a\right) \frac{j^2t}{\sqrt{i^2+(jt)^2}}+pa=0.
\end{equation}
Eliminate variable $p$ from Eq.~(\ref{correction_2}) and Eq.~(\ref{correction_3}), we get:
\begin{equation}
\label{correction_4}
\sum_{(i, j) \neq (0, 0)} u_2'\left (\sqrt{i^2+(jt)^2}a\right) \frac{i^2-(jt)^2}{\sqrt{i^2+(jt)^2}}=0.
\end{equation}
Integrating Eq.~(\ref{correction_4}) over $a$ will simplify it and gives:
\begin{equation}
\label{correction_5}
\sum_{(i, j) \neq (0, 0)} u_2\left (\sqrt{i^2+(jt)^2}a\right) \frac{i^2-(jt)^2}{i^2+(jt)^2}=C, 
\end{equation}
where $C$ is an arbitrary constant. Eq.~(\ref{correction_5}) is a necessary condition for stability. Generally, a potential function does not satisfy this condition over a range of $a$. However, for any potential function $u_2^0(r)$, let:
\begin{equation}
\label{correction_6}
u_2^1(r)=-\frac{1}{2} \sum_{(i, j) \neq (0, 0)}u_2^0\left(r \sqrt{i^2+(jt)^2}\right)\frac{i^2-(jt)^2}{i^2+(jt)^2}+C \mbox{ }(0.9<r<1.1),
\end{equation}
Then, the potential $u_2(r)=u_2^0(r)+u_2^1(r)$ satisfies Eq.~(\ref{correction_5}) over the range $0.9<a<1.1$. 
Constant $C$ in Eq.~(\ref{correction_6}) is chosen so that $u_2^1(1)=0$. The ``correction'' $u_2^1(r)$ is usually much smaller than $u_2^0(r)$.

We have applied this correction to our higher-order potential for the rectangular lattice with aspect ratio $b/a=2$ (Eq.~(\ref{Rectangle_H_7_Eq})). 
After that, we do simulated annealing using the corrected potential at different pressures. 
We found that the rectangular lattice with aspect ratio $b/a=2$ is indeed the ground state of the corrected potential over the pressure range $0.98<p<1.87$.

\end{document}